# Termodinámica Estadística de macromoléculas sobre membranas biológicas

Luis Silva-Acosta[1], M. G. González-Asevedo[2] y R. Juárez-Maldonado[1,3]*

[1] Unidad Académica de Física, Universidad Autónoma de Zacatecas, Calzada Solidaridad Esquina con Paseo la Bufa S/N, C. P. 98060 Zacatecas, Zac., México.

[2] Instituto de Física "Manuel Sandoval Vallarta", Universidad Autónoma de San Luis Potosí, San Luis Potosí, San Luis Potosí, México.

[3] Facultad de Ciencias Exactas, Universidad Juárez Del Estado de Durango, Avenida Veterinaria #210, colonia Valle del Sur, C. P. 34120 Durango, Durango, México

*Email: kblogw@gmail.com

## *Resumen*

En este artículo se desarrolla explícitamente el formalismo termodinámico estadístico general para el estudio de sistemas restringidos a superficies de revolución, particularmente se deriva la ecuación de estado del virial para el caso de la esfera y el toro. Se construye un modelo para el estudio de las propiedades termodinámicas de macromoléculas sobre membranas biológicas.

## *Abstract*

In this paper we develop explicitly the general Statistical Thermodynamics formalism to study of systems restricted to lie on revolution surfaces, particularly we derive the virial state equation of sphere and torus. A model to study the termodynamical properties of macromolecules on biological membranes is constructed.

## *I. Introducción*

La gran mayoría de los fenómenos que ocurren en la naturaleza se manifiestan bajo la influencia de algún mecanismo que los restringe a vivir en un espacio limitado. Tal es el caso por ejemplo de los organelos celulares los cuales se encuentran confinados en el interior de la membrana celular o las diferentes biomoléculas que se encuentran restringidas a moverse sobre la pared celular [1,2]. Otro ejemplo es toda clase de moléculas, seres microscópicos o partículas suspendidas en interfaces [3] (agua-aire, aceite-agua, burbujas, etc.). La dinámica misma de la vida está restringida a la superficie del planeta, así como los huracanes y corrientes de aire, entre muchos ejemplos más.

Resulta pues de fundamental importancia investigar los efectos que pueda tener por ejemplo la curvatura sobre las propiedades dinámicas y estáticas de dichos sistemas. Se han hecho algunos intentos por estudiar las propiedades termodinámicas de sistemas restringidos a superficies esféricas [4] y cilíndricas [5], también se han hecho estudios de como la curvatura afecta las propiedades de auto-difusión [6]. El Formalismo de la Mecánica Estadística es lo suficientemente general para incluir tales restricciones y efectos, así pues existen determinadas sutilezas a la hora de los cálculos explícitos de propiedades tales como la presión o como las funciones de distribución, sobre todo cuando se incluyen los efectos de tamaño finito o de frontera.

El propósito del presente trabajo es desarrollar explícitamente la estructura metodológica formal de la Mecánica Estadística Clásica para el caso particular de sistemas restringidos a vivir sobre superficies de revolución. Una versión más general que incluye el caso de sistemas multicomponente sobre variedades Riemannianas de dimensión arbitraria está siendo preparada para su publicación [7]. Y una vez construido ese marco metodológico formal, aplicarlo al caso particular de macromoléculas que viven sobre la pared celular.

El artículo se organiza de la siguiente forma. En la sección II se plantea el formalismo canónico (Hamiltoniano) para la dinámica de sistemas restringidos a superficies de revolución. En la sección III se revisan los conceptos de Mecánica Estadística bajo la teoría de los ensambles: Micro-canónico, Canónico y Gran-Canónico. En la sección IV se analiza el caso particular de gases ideales, además de derivar la fórmula barométrica. Finalmente en la sección V se da un sumario de los principales resultados del presente trabajo.

## *II. Formulación Hamiltoniana sobre superficies de revolución*

Considere una partícula de masa *m* restringida a moverse sobre la superficie de revolución definida por $\boldsymbol{S}(u,\phi) = \big(f(u)\cos\phi, f(u)\sin\phi, g(u)\big)$ (ver Apéndice A y Ref. [8]) y bajo la influencia del campo externo $V = V(u,\phi)$. La dinámica de dicha partícula está completamente contenida en las ecuaciones de Hamilton [9]

$$\dot{p}_u = -\frac{\partial \boldsymbol{H}}{\partial u}$$

$$\dot{p}_\phi = -\frac{\partial \boldsymbol{H}}{\partial \phi}$$

$$\dot{u} = -\frac{\partial \boldsymbol{H}}{\partial p_u}$$

$$\dot{\phi} = -\frac{\partial H}{\partial p_\phi}$$

Donde $p_u = \frac{\partial L}{\partial \dot{u}}$, $p_\phi = \frac{\partial L}{\partial \dot{\phi}}$, siendo $L$ la función Lagrangiana

$$L = \frac{1}{2}m\left\{\left[\left(\frac{\partial f}{\partial u}\right)^2 + \left(\frac{\partial g}{\partial u}\right)^2\right]\dot{u}^2 + [f(u)]^2\dot{\phi}^2\right\} - V(u,\phi),$$

y $H$ el Hamiltoniano

$$H = \frac{p_u^2}{2m\left[\left(\frac{df}{du}\right)^2 + \left(\frac{dg}{d\phi}\right)^2\right]} + \frac{p_\phi^2}{2m[f(u)]^2} + V(u,\phi).$$

Por ejemplo, para el caso particular de la esfera de radio R, $f(u) = R\sin u$ y $g(u) = R\cos u$; por lo que el el Hamiltoniano es simplemente

$$H = \frac{p_u^2}{2mR^2} + \frac{p_\phi^2}{2mR^2\sin^2 u} + V(u,\phi).$$

## III. Ensambles Estadísticos

Consideremos un sistema de $N$ partículas de masa $m$ restringido a la superficie de revolución $\mathbf{S}(u,\varphi) = \big(f(u)\cos\phi, f(u)\sin\phi, g(u)\big)$. Si denotamos un punto del espacio fase de este sistema por $(\mathbf{q},\mathbf{p}) = (q_1,\ldots q_N, p_1,\ldots p_N) = (u_1,\phi_1,\ldots u_N,\phi_N, p_{u_1}, p_{\phi_1},\ldots p_{u_N}, p_{\phi_N})$, el potencial de interacción entre pares de partículas con $U_{ij} = U(\mathbf{q_i} - \mathbf{q_j})$ y un campo externo con $V_i = V(\mathbf{q_i})$, entonces el Hamiltoniano del sistema está dado por

$$H_N(\mathbf{q},\mathbf{p}) = \frac{1}{2m}\sum_{i=1}^{N}\frac{p_{u_i}^2}{\left[\left(\frac{df}{du}\right)^2 + \left(\frac{dg}{d\phi}\right)^2\right]} + \frac{p_{\phi_i}^2}{[f(u)]^2} + U_N(u,\phi) + V_N(u,\phi),$$

Donde $U_N(u,\phi) = \sum_{i<j}^{N} U_{ij}$ y $V_N(u,\phi) = \sum_{i=1}^{N} V_i$.

### Ensamble Micro-canónico

En este ensamble el sistema está completamente aislado, las variables características son el número de partículas $N$, el volumen $A$ (que en el caso de superficies, se refiere al área de interés), y la energía $E$; que al mantenerse fijos determinan de manera única el volumen accesible $\Omega(N,V,E)$ del espacio fase al sistema. Este volumen accesible es simplemente el volumen de la hipersuperficie de energía $H_N(\mathbf{q},\mathbf{p}) = E$:

$$\Omega(N,V,E) = \int_{H(q,p)=E} d^N q \, d^N p$$

Donde $d^N q d^N p = dq_1 dq_2 \ldots dq_N dp_1 dp_2 \ldots dp_N$. En principio, dado $\Omega(N,V,E)$, todas las propiedades termodinámicas del sistema pueden ser calculadas a partir de la expresión para la entropía

$$S = k \ln \Omega(N, V, E),$$

mediante las ecuaciones de estado

$$\frac{1}{T} = \left(\frac{\partial S}{\partial E}\right)_{V,N}$$

$$\frac{p}{T} = \left(\frac{\partial S}{\partial V}\right)_{E,N}$$

$$-\frac{\mu}{T} = \left(\frac{\partial S}{\partial N}\right)_{E,V}$$

siendo $k$ la constante de Boltzmann, $T$ la temperatura, $p$ la presión y $\mu$ el potencial químico. Sin embargo, el cálculo de la integral en general no es simple.

Aún así, trabajaremos la integral hasta llevarla a una forma convencionalmente. Sea $\mathbf{s}(w_1, \ldots, w_{N-1}) = \left(s_1(w_1, \ldots, w_{N-1}), \ldots, s_{N-1}(w_1, \ldots, w_{N-1})\right)$ la forma parametrizada de la hipersuperficie de energía $H(\mathbf{q},\mathbf{p})=E$, entonces el área $\Sigma(E)$ de esta hipersuperficie corresponde con el volumen accesible del sistema $\Omega(N,V,E)$. Consecuentemente, otra forma de expresar el volumen accesible es

$$\Omega(N, V, E) = \int_{\Sigma(E)} \sqrt{g}\, d^{N-1}w$$

donde $g$ es el determinante del tensor métrico $g_{ij} = \frac{\partial \mathbf{s}}{\partial w_i} \cdot \frac{\partial \mathbf{s}}{\partial w_j}$ de la variedad Riemanniana [ver apéndice B] de fase definida por la hipersuperficie de energía, y $d^{N-1}w = dw_1 dw_2 \ldots dw_{N-1}$. Esta última expresión para el volumen accesible es lo bastante general y lo suficientemente simple para evaluar algunos casos idealizados.

## b). Ensamble Canónico

En este ensamble el sistema intercambia energía térmica con el medio ambiente en el cual se encuentra, pero no materia. Es decir, está en equilibrio con un reservorio térmico a temperatura $T$, con el cual puede intercambiar calor pero no trabajo y su número de partículas $N$ es constante. Las variables características son el número de partículas $N$, el volumen $V$ y la temperatura $T$. Dado que el sistema no está aislado, su energía no es constante y fluctúa (puede intercambiar energía con el reservorio), entonces, sólo podemos hablar de probabilidad de que el sistema adopte una energía determinada para un dado valor de la temperatura del reservorio. La probabilidad buscada es

$$P_E = \frac{e^{-\beta E}}{Z_N}.$$

Donde la función de partición canónica $Z_N$ es una constante de normalización impuesta para que la suma de las probabilidades de todos los estados sea uno. Se define para estos sistemas en consideración como

$$Z_N = \frac{1}{N! h^{2N}} \int d^N q \, d^N p \, e^{-\beta H_N(q,p)},$$

donde $H$ el Hamiltoniano del sistema, $N$ el número de partículas, $\beta = \frac{1}{kT}$ y, $k$ y $h$ las constantes de Boltzmann y Plank respectivamente. Esta ecuación nos da la energía libre de Helmholtz $F = -kT \ln Z_N$ del sistema en función de sus variables naturales, lo que supone un conocimiento termodinámico exhaustivo del sistema mediante las ecuaciones de estado.

$$p = -\left(\frac{\partial F}{\partial A}\right)_{N,T}$$

$$S = -\left(\frac{\partial F}{\partial T}\right)_{N,A}$$

$$\mu = \frac{\partial F}{\partial N}$$

Por tanto conocer la función de partición es resolver el problema estadístico. Trabajaremos entonces con esta función de partición al igual que en el inciso a) hasta llevarla a una forma convenientemente simple. Consideremos el sistema de $N$ partículas que se describe al inicio de la sección III. Tomando el Hamiltoniano para este sistema, la función de partición canónica es

$$Z_N = \frac{1}{N! h^{2N}} \int d^N u \, d^N \phi \, e^{-\beta(U_N(q)+V_N(q))} \int d^N p_u \, d^N p_\phi \, e^{-\left(\frac{\beta}{2m}\right)\Sigma_i\left\{\left[\left(\frac{df}{du}\right)^2+\left(\frac{dg}{du}\right)^2\right]^{-1} p_{u_i}^2 + [f(u)]^2 p_{\phi_i}^2\right\}}$$

Integrando sobre todos los momentos tenemos

$$Z_N = \frac{1}{N! \Lambda^{2N}} \int \prod_{i=1}^{N} dA_i \, e^{-\beta(U_N(u,\phi)+V_N(u,\phi))}$$

Donde $dA_i = \sqrt{\left[\left(\frac{df}{du_i}\right)^2 + \left(\frac{dg}{du_i}\right)^2\right][f(u_i)]^2} du_i d\phi_i$ y $\Lambda = \frac{h}{\sqrt{2\pi mkT}}$ es la longitud de onda térmica de de Broglie. Nuevamente, esta última expresión para la función de partición canónica es lo bastante general y suficientemente simple para evaluar algunos casos de interés.

## c). Ensamble Gran-canónico

En este ensamble el sistema intercambia materia y energía con el medio ambiente en el cual se encuentra inserto. El medio ambiente actúa entonces a la vez como reservorio de temperatura y de partículas. Las variables características son el potencial químico $\mu$, el volumen $V$ y la temperatura $T$; Vamos a suponer que el volumen del sistema se mantiene constante, de manera que pensamos que el sistema se encuentra separado del reservorio por paredes rígidas pero permeables. Si el sistema se encuentra en equilibrio con el reservorio, el número de partículas no será mas constante, pero podemos asumir que tanto su energía media como su densidad media de partículas, permanecen constantes, de esta forma los estados accesibles del sistema son en este caso los autoestados de la energía para una partícula, para dos partículas, etc..

Al igual que en el ensamble canónico, a partir de la gran función de partición se pueden calcular expresiones para los valores esperados o promedios de las funciones de estado y se define para estos sistemas en consideración como

$$\Xi(\mu, A, T) = \sum_{N \geq 0} \lambda^N Z_N(A, T)$$

Donde $\lambda = e^{\beta\mu}$. Esta ecuación nos lleva al gran potencial que esta dado por $\Omega = -pA = -kT \ln \Xi$, entonces es fácil demostrar que el número promedio de partículas es

$$\bar{N} = kT \left(\frac{\partial \ln \Xi}{\partial \mu}\right)_{A,T}$$

$$S = k \ln \Xi + kT \left(\frac{\partial \ln \Xi}{\partial T}\right)_{V,\mu}$$

## *IV. Gases Ideales*

### a). Ensamble Micro-canónico

Para este caso de gases ideales, se puede resolver la integral del volumen accesible con cualquiera de las dos formas descritas antes. Tomando la primera de las formas tenemos

$$\Omega(N, A, E) = \int_{\Sigma_{i=1}^N \frac{p_{u_i}^2}{A(u)} + \frac{p_{\phi_i}^2}{B(u)} = 2mE} d^N u \, d^N \phi \, d^N p_u \, d^N p_\phi$$

Donde $A(u) = \left[\left(\frac{df}{du}\right)^2 + \left(\frac{dg}{du}\right)^2\right]$ y $B(u) = [f(u)]^2$. Para resolver esta integral, hay que tener en cuenta que $A(u)$ y $B(u)$ no dependen de los momentos, por tanto podemos hacer un cambio de variables $\left\{x_1^2 = \frac{p_{u_1}^2}{A(u_1)}, \ldots, x_N^2 = \frac{p_{u_N}^2}{A(u_N)}, x_{N+1}^2 = \frac{p_{\phi_1}^2}{B(u_1)}, \ldots, x_{2N}^2 = \frac{p_{\phi_N}^2}{B(u_N)}\right\}$. Entonces la integral toma la forma

$$\Omega(N, A, E) = \int dA_1 \cdots dA_N \int_{\Sigma_{i=1}^{2N} x_i^2 = R^2} dx_1 \cdots dx_{2N}$$

Donde $dA_j = \sqrt{A(u_j)B(u_j)} du_j d\phi_j$ es la superficie del elemento j-esimo y $R = \sqrt{2mE}$, Entonces el problema se reduce a calcular el área de la hipersuperficie $S_n(R)$ de radio R y dimensión n=2N. La solución es $\frac{2\pi^{\frac{n}{2}}}{\Gamma(\frac{n}{2})} R^{n-1}$ (ver apéndice), donde $\Gamma$ es la función gamma. Finalmente, se pueden obtener todas las propiedades del sistema, a partir de la definición de la entropía la cual puede escribirse como

$$S = k \ln \left( \frac{2\pi^N}{\Gamma(N)} (2mE)^{N-\frac{1}{2}} A^N \right)$$

Donde $A = \int dA = \int \sqrt{A(u)B(u)}\, dud\phi$ es el área de la superficie de revolución o cualquier región de interés sobre dicha superficie.

## b). Ensamble Canónico

Recordemos que la función de partición canónica es

$$Z_N = \frac{1}{N!\, h^{2N}} \int \prod_{i=1}^{N} dA_i \, e^{-\beta(U(u,\phi)+V(u,\phi))}$$

Entonces para el caso de un gas ideal y en ausencia de un campo externo, la función de partición es simplemente

$$Z_N = \frac{A^N}{N!\, h^{2N}}$$

Donde $A = \int dA = \int \sqrt{\left[\left(\frac{df}{du}\right)^2 + \left(\frac{dg}{du}\right)^2\right][f(u)]^2}\, dud\phi$, es el área de la superficie de revolución o cualquier región de interés. Por ejemplo para la esfera $A = 4\pi R^2$. De aquí es trivial obtener las ecuaciones de estado a partir de la energía libre de Helmholtz. En particular, la ecuación de la presión es

$$p = -\left(\frac{\partial F}{\partial A}\right)_{N,T} = \frac{NkT}{A}$$

La cual coincide con la ecuación de estado de la presión para el gas ideal contenido en un volumen V, solo que esta vez, el gas está sobre la superficie.

## c). Ensamble Gran-canónico

Para este ensamble, hay que notar que la función de partición canónica para un gas ideal es

$$Z_N = \frac{1}{N!} Z_1^N$$

Donde $Z_1 = \frac{1}{h^2} \int dA e^{-\beta V(u,\phi)}$ es la función de partición para una sola partícula bajo la influencia de un campo externo. En ausencia de un campo externo, la función de partición canónica del sistema es simplemente

$$Z_N = \frac{A^N}{N!\,h^{2N}} = \frac{(2\pi mkT)^N}{N!\,h^{2N}} A^N$$

como ya se había visto. Entonces, de aquí la gran función de partición es

$$\Xi(\mu, A, T) = \sum_{N=1}^{\infty} \frac{\left(\frac{\lambda A}{h^2}\right)^N}{N!} = e^{\frac{\lambda A}{h^2}}$$

Por tanto, tenemos el número promedio de partículas como

$$\bar{N} = kT\left(\frac{\partial \ln \Xi}{\partial \mu}\right)_{A,T} = \frac{\lambda A}{h^2}$$

Ahora si sistema está bajo la influencia de un campo externo, y si denotamos la densidad de partículas por $\rho(q)$ entonces el número promedio de partículas en la superficie es

$$\bar{N} = \int dA\, \rho(u, \phi)$$

Por otra parte, la gran función de partición es

$$\Xi(\mu, A, T) = \sum_{N=1}^{\infty} \frac{(\lambda Z_1)^N}{N!} = e^{\lambda Z_1}$$

entonces

$$\bar{N} = \lambda \frac{\partial \ln \Xi}{\partial \lambda} = \lambda Z_1$$

Y por tanto, igualando estas dos ecuaciones obtenemos la formula barométrica.

$$\rho(u, \phi) = \rho_0 e^{-\beta V(u,\phi)},$$

donde $\rho_0 = \frac{(2\pi mkT)}{h^2} e^{\beta \mu}$. Esta fórmula es importante porque es una primera aproximación al estudio de la estructura de sistemas inhomogeneos, que en este caso la inhomogeneidad es provocada por la presencia de un campo externo. También describe en una primera versión la respuesta de un sistema ante perturbaciones externas, lo cual puede ser de gran utilidad para el estudio de las propiedades estructurales de dominios sobre la membrana celular o deformaciones bajo la presencia de capos externos [Ref. Pendientes].

## *V. Sistemas interactuantes vía expansión de virial*

En esta sección estamos interesados en el caso particular donde el sistema de N partículas es libre de campos externos $V = 0$, pero existe un potencial de interacción a pares entre las partículas del sistema, de tal forma que la energía potencial del sistema sea $U_N = \sum_{\{i<j\}} U(q_i - q_j)$. Para nuestro estudio usaremos el formalismo del ensamble canónico, ya que, mediante este ensamble se pueden incluir los efectos de tamaño finito que pueda tener el sistema (los cual es importante en el caso de sistemas biológicos, como macromoléculas sobre la pared celular). Ente este caso, la función de partición canónica es de la forma

$$Z_N = \frac{1}{N!\,\Lambda^{2N}} \int \prod_{i=1}^{N} dA_i \, e^{-\beta U_N(\boldsymbol{u},\boldsymbol{\phi})}$$

Como ya vimos en la sección (IIIb) la ecuación de la presión es simplemente

$$p = -\left(\frac{\partial F}{\partial A}\right)_{N,T} = kT\left(\frac{\partial \ln Z_N}{\partial A}\right)_{N,T}$$

sin embargo, la integral que aparece en $Z_N$ salvo en gases ideales, en la práctica es imposible de calcular. Una ruta alternativa, es hacer una expansión en serie de potencias de la densidad, de tal manera que la ecuación de la presión queda de la forma

$$\frac{p}{kT} = \rho\left(1 + \sum_{i=2}^{\infty} B_i(N,A)\rho^{i-1}\right) = \rho(1 + B_2(N,A)\rho + B_3(N,A) + \cdots)$$

la cual es conocida como expansión del virial y los coeficientes $B_k(N,A)$ son llamados coeficientes del virial. Nótese que los coeficientes en general de penden, separadamente, tanto del número de partículas como del volumen del sistema. Se puede demostrar que la forma general de la dependencia en el volumen para los coeficientes del viarial [Lebowitz] es

$$B_k(A) = B'_k(A) - \frac{A}{k-1}\frac{\partial B'_k(A)}{\partial A}$$

donde $B'_k(A)$ son las conocidas integrales de Mayer [Hill]

$$B'_k(A) = -\frac{1}{(n-1)A}\int \cdots \int S_{12\ldots k}\, dA_1 dA_2 \ldots dA_k$$

siendo $f_{ij} = e^{-\beta U(\boldsymbol{q}_i - \boldsymbol{q}_j)} - 1$ las funciones de Mayer. En este artículo nos limitaremos al cálculo del segundo coeficiente del virial, $B_2(N,A)$, incluyendo los efectos de volumen y número de partículas finito toma la forma [lebowitz]

$$B_2(N,A) = \left(1 - \frac{1}{N}\right)\left(B'_k(A) - \frac{A}{k-1}\frac{\partial B'_k(A)}{\partial A}\right)$$

con

$$B'_2(A) = -\frac{1}{2A}\iint dA_1 dA_2 f_{12} = -\frac{1}{2A}\iint dA_1 dA_2 (e^{-\beta U(\boldsymbol{q}_1,\boldsymbol{q}_2)} - 1)$$

Note que en general, el potencial depende tanto de la posición de la partícula **1** como de la **2**. Sin embargo, un caso de especial interés es cuando solo depende de la distancia geodésica[1] $r$ entre ambas, en dado caso

---

[1] En este trabajo sólo discutiremos sistemas homogéneos e isotrópicos, por lo que la interacción solo depende de la distancia geodésica entre partículas.

$$B'_2(A) = -\frac{1}{2A} \iint dA dA_r \left(e^{-\beta U(r)} - 1\right) = -\frac{1}{2} \int dA_r \left(e^{-\beta U(r)} - 1\right)$$

sonde $dA_r = \sqrt{\left[\left(\frac{df}{du}\right)^2 + \left(\frac{dg}{du}\right)^2\right]} [f(u)]^2 du d\phi$.

### a). Sistemas sobre una superficie esférica: Membrana celuar esferoidal

En esta sección estudiaremos en detalle la ecuación de estado de un sistema de partículas restringido a una superficie esférica, este sistema es un buen modelo de sistemas de macromoléculas sobre la membrana celular.

En este caso $dA_r = R^2 \sin u$, y si consideramos una interacción tipo esfera dura, cuyo potencial es

$$U(uR) = \begin{cases} \infty & si \ uR < \sigma \\ 0 & si \ uR \geq \sigma \end{cases}$$

donde $\sigma$ es el diámetro geodésico de una partícula sobre la superficie esférica y $uR$ es la distancia geodésica entre partículas. Este potencial de interacción es una primera aproximación a la interacción entre macromoléculas sobre la pared celular, la cual en el limite de bajas densidades puede decirse que es buena. La integral que aparece al calcular $B'(A)$ es muy simple, el resultado es

$$B_2'(A) = 2\pi R^2 \sin^2\left(\frac{\sigma}{2R}\right) = \frac{A}{2} \sin^2\left(\frac{\sigma}{\sqrt{A/\pi}}\right)$$

luego el segundo coeficiente del virial es

$$B_2(N, A) = \left(1 - \frac{1}{N}\right) \frac{A}{4} \frac{\sigma}{R} \sin\left(\frac{\sigma}{2R}\right)$$

y finalmente la ecuación de estado del sistema es

$$\frac{p}{kT} = \rho \left\{ 1 + \left(1 - \frac{1}{N}\right) \frac{A}{4} \frac{\sigma}{R} \sin\left(\frac{\sigma}{2R}\right) \rho \right\}$$

Solo graficamos N=10, para otros valores no hay cambios significativos. Observe que a partir de R=3, todas las curvas son una misma, lo cual indica que a partir de cierta asimetría entre los tamaños de las macromoléculas y la célula el comportamiento será siempre el mismo.

### b). Sistemas sobre una superficie toroidal: Eritrocito

En este caso, el segundo coeficiente del virial toma la forma:

$$B_2(N, A) = \left(1 - \frac{1}{N}\right) \left[\frac{\sigma^2}{2} + \sigma^2 \frac{r}{R} \cos\left(\frac{\sigma}{r}\right)\right]$$

y finalmente la ecuación de estado de un sistema partículas duras sobre la superficie del toro es

$$\frac{p}{kT} = \rho(1 + B_2(N,A)\rho + \cdots) \approx \rho\left\{1 + \left(1 - \frac{1}{N}\right)\left[\frac{\sigma^2}{2} + \sigma^2 \frac{r}{R}\cos\left(\frac{\sigma}{r}\right)\right]\rho\right\}$$

Específicamente en la figura se grafican, para varios radios de R, observe que conforme R crece la ecuación de estado se satura hasta llegar, como es natural, al caso de un sistema bidimensional plano.

## *Apéndice A: Superficies de revolución*

Una *superficie de revolución* es una superficie generada al rotar una curva plana en torno a una recta contenida en ese mismo plano.

Si *h(x,z)=0* es la ecuación de una curva sobre el plano *XZ* y gira entorno al eje *Z*, las sustituciones necesarias para generar la superficie de revolución son

$$z \to z \quad x \to \sqrt{x^2 + y^2},$$

de tal manera que la ecuación de la superficie es

$$F\left(\sqrt{x^2 + y^2}, z\right) = 0.$$

Equivalentemente si la curva esta dada por la trayectoria parametrizada *γ=(f(u),0,g(u))*, entonces la superficie de revolución, al girarla alrededor del eje *Z*, es simplemente

$$s(u, \phi) = (f(u)\cos\phi, f(u)\sin\phi, g(u)),$$

donde ϕ es el ángulo de revolución (Ver Fig. A.1).

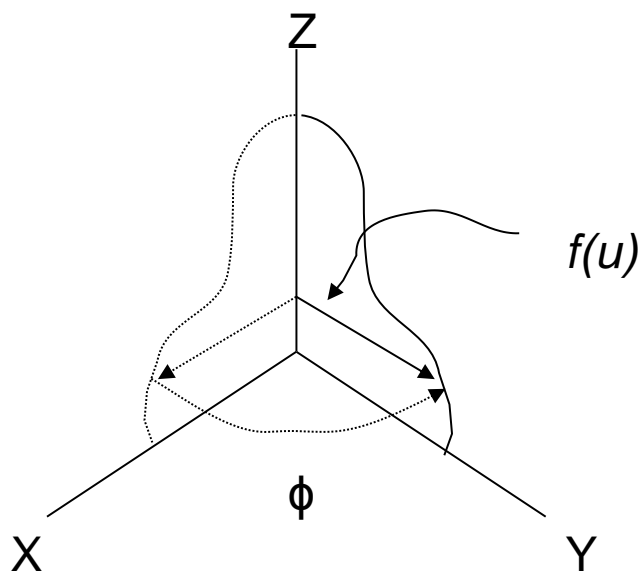

*Fig. A.1: Superficie de Revolución*

A continuación enumeramos algunos ejemplos particulares de interés:

> Esfera: *f(u) = R sin(u), g(u) = R cos(u).*
>
> Toro: *f(u) = ( R+r cos (u) ), g(u) = r sin(u).*
>
> Hiperboloide: *f(u) = a cosh(u), g(u) = b cosh(u).*
>
> Paraboloide: *f(u) = a u².*

## *Apéndice B: Elementos de Geometría Riemanniana*

Para nosotros una variedad Riemanniana de dimensión $d$ es un subconjunto de $\mathcal{R}^n$ parametrizable mediante el mapeo (diferenciable cuantas veces sea necesario) $\boldsymbol{x}(w_1, \ldots, w_d) = \left(x_1(w_1, \ldots, w_d), \ldots, x_n(w_1, \ldots, w_n)\right)$ con $x_k, w_k \in \mathcal{R}$. Además de estar dotado con un producto interno, tal que, si $\boldsymbol{a}$ y $\boldsymbol{b}$ son dos vectores tangentes a la variedad su producto se define a través del tensor métrico $g_{ij} = \frac{\partial \boldsymbol{x}}{\partial w_i} \cdot \frac{\partial \boldsymbol{x}}{\partial w_j}$ como $\boldsymbol{a} \cdot \boldsymbol{b} = a^i g_{ij} b^j$. El elemento de longitud sobre la variedad esta dado por $ds^2 = g_{ij} dx^i dx^j$, similarmente, el elemento de volumen viene dado por $d\Sigma = \sqrt{\text{g}} dw_1 \ldots dw_d$ donde $\text{g} = \det(g)$ es el determinante del tensor métrico. (Para más rigurosa introducción a la geometría Riemanniana ver Ref. 10)

## *Bibliografía:*